\documentclass[preprint,aps,superscriptaddress]{revtex4}
\usepackage{graphicx}
\usepackage{amsmath}
\usepackage{color}
\usepackage{amssymb}
\newcommand{\be}{\begin{equation}}
\newcommand{\ee}{\end{equation}}
\newcommand{\bea}{\begin{eqnarray}}
\newcommand{\eea}{\end{eqnarray}}
\newcommand{\pr}{\partial}
\newcommand{\nno}{\nonumber}
\newcommand{\bse}{\begin{subequations}}
\newcommand{\ese}{\end{subequations}}

\def\gsim{\lower0.5ex\hbox{$\:\buildrel >\over\sim\:$}}
\def\lsim{\lower0.5ex\hbox{$\:\buildrel <\over\sim\:$}}

\begin{document}
\title{Constraining PCP Violating Varying Alpha Theory Through Laboratory
Experiments}
\author{Debaprasad Maity
\footnote{debu.imsc@gmail.com}}
\affiliation{Department of Physics and Center for
Theoretical Sciences, National Taiwan
University, Taipei 10617, Taiwan}
\affiliation{Leung Center for Cosmology and Particle Astrophysics\\
National Taiwan University, Taipei 106, Taiwan}
\author{Pisin Chen\footnote{chen@slac.stanford.edu}}
\affiliation{Department of Physics and Center for 
Theoretical Sciences, National Taiwan
University, Taipei 10617, Taiwan}
\affiliation{Leung Center for Cosmology and Particle Astrophysics\\
National Taiwan University, Taipei 106, Taiwan}
\affiliation{Kavli Institute for Particle Astrophysics and Cosmology\\
SLAC National Accelerator Laboratory, Menlo Park, CA 94025, U.S.A.}

\begin{abstract}
In this report we have studied the implication of 
a parity and charge-parity (PCP) violating interaction 
in varying alpha theory. Due to this interaction,     
the state of photon polarization can change when it passes
through a strong background magnetic field. 
We have calculated the optical rotation and ellipticity of the plane
of polarization of an electromagnetic wave and tested our results against 
different laboratory experiments. Our model contains a PCP
violating parameter $\beta$ and a scale of alpha variation $\omega$.
By analyzing the laboratory experimental data, we found the most stringent
constraints on our model parameters to be $1\leq \omega \leq 10^{13}~
\mbox{GeV}^2$ and $-0.5\leq \beta \leq 0.5$. We also found that
with the existing experimental input parameters it is very
difficult to detect the ellipticity in the near future.    
\end{abstract}

\maketitle

\section{Introduction}\label{intro}
Parity violation 
is one of the simplest straightforward extension of the standard model
of physics. So far, violation of parity (P) and charge-parity (CP) 
has been observed only in the electroweak sector of the
standard model particle physics. Considering this as a 
guiding principle, in recent years 
several models of PCP violation have been constructed 
\cite{carroll,marc,contaldi,soda,debu}.
The basic idea of all these models is to add an explicit  
parity violating term in the Lagrangian. Because of its
nature, this parity violating term leads to cosmic 
birefringence \cite{carroll,marc} and left-right asymmetry
in the gravitational wave dynamics \cite{contaldi,soda}. 
String theory inspired models with non-standard parity-violating interactions 
have also been discussed \cite{debu}. 
Recently we have constructed a parity and 
charge-parity (PCP) violating model \cite{debupisin}
in the framework of ``varying alpha theory". 
Some aspects of our model are similar to that 
proposed by Carroll \cite{carroll}.
But as we have argued, our model has the 
advantage over that of Carroll's in that 
the origin of the parity violation may be more physically motivated. 

Sting theory has given us sufficient theoretical
motivation to consider theories of varying fundamental constants in 
nature. As is well known, string theory is fundamentally
a higher dimensional theory.
In principle, therefore, all the so-called fundamental constants 
in our four dimensional world 
may actually be spacetime dependent as a result of the dimensional reduction. 
Meanwhile, increasingly high 
precession cosmological as well as laboratory experiments give us 
hope that signature of new physics, including those that give 
rise to variation of fundamental constants, may 
emerge in the near future.    
  
A consistent, gauge invariant and Lorentz invariant 
framework of $\alpha$
variability was first proposed by Bekenstein \cite{bek2}.
Subsequently this subject has been studied quite extensively  
from the theoretical side \cite{bsm,bsm1,barrow}
as well as from the observational side 
\cite{murphy,webb,webb2,dent,landau}.

Apart from verifying this notion from cosmological/astrophysical and high energy
collider experiments, it is also important to study
various purely laboratory based experiments which can
provide complementary results. In this report we
focus on its connection to a particular class of laboratory experiments which make use of the
conversion of axion or any other low-mass
(pseudo)scalar particle into photon in the presence of an
electromagnetic field. These include the
Brookhaven-Fermilab-Rutherford-Trieste (BFRT) experiment
\cite{bfrt}, the Italian PVLAS experiment
\cite{pvlas} and several other experiments such as
Q\&A \cite{qa}, BMV \cite{bmv}, etc., which are either
already in progress or ready to be built.
All these experiments are expected to produce
(pseudo)scalars from polarized laser beams, which
are allowed to propagate in a transverse, constant and
homogeneous magnetic field. In addition to the direct production of (pseudo)scalar particles, modification of the polarization of
light can in principle be induced by
its coupling with pseudo-scalar axions as it propagates through a transverse magnetic field
\cite{raffelt,zavattini,pdpj,birch,kendall,cf,cfj,ralston}.
The model that we recently introduced also exhibits this effect 
induced from the PCP violating term in our varying fine structure 
constant theory \cite{debupisin}.
This motivates us to use a different class of experiments
to constrain the parameters of a given
varying fine structure constant theory. Such approach
has not been explored before. So far varying alpha theory
has been constrained mostly based upon the observations on the possible
variation of fine structure constant. As we mentioned therefore by introducing 
PCP violation in a varying alpha theory, we actually
unified different class of experimental observation
in a single framework.
The birefringence and the dichroism of the 
vacuum induced from the non-trivial coupling of photon and the
varying alpha scalar can be tested in those class of experiments, 
which in turn will constrain the parameter space of a given
varying alpha model.
In this paper we aim to address these issues and test it against
above mentioned various laboratory based experiments to constrain our
model parameters.

We organize this paper as follows: in Section \ref{sec1}, we 
review our PCP violating ``varying alpha theory" constructed in 
\cite{debupisin}. 
In next section \ref{sec2}, we will calculate
the effect of background magnetic field on the 
plane of polarization of electromagnetic field. 
As we have mentioned before, we will particularly focus 
on various laboratory based experiments which 
measure the rotation of the plane of polarization and 
the ellipticity of electromagnetic field 
through scalar-photon conversion mechanism in the strong 
background magnetic field. We will analytically calculate
the expression for rotation and ellipticity.   
Then in section \ref{sec3}, we will use 
the laboratory based experimental bounds on
rotation and ellipticity to constrain our model
parameter. 
Concluding remarks and future prospects are provided in 
Section \ref{con}.

\section{Parity violating varying-alpha theory}\label{sec1}
In this section we will review our previous construction of 
PCP violating varying alpha theory \cite{debupisin}. 
A varying alpha theory \cite{bek2,bsm,bsm1} 
is usually referred to as a theory of spacetime variation of the electric charge of any matter field,
parameterized by $e=e_{0} e^{\phi(x)}$, where 
$e_0$  denotes the coupling constant and $\phi(x)$ 
is a dimensionless scalar field. 
The fine-structure constant in such a theory is therefore $\alpha = e_0^2 e^{2\phi(x)}$. 
This theory has been constructed based upon the shift 
symmetry in $\phi$ i.e. $\phi\rightarrow \phi + c$ and
the modified U(1) gauge transformation 
$e^\phi A_\mu \rightarrow e^\phi A_\mu +\chi_{,\mu }.$
From the above symmetry considerations, the unique gauge-invariant 
and shift-symmetric Lagrangian for the modified electromagnetic field 
and the scalar field can be written as 
\bea
S= - \frac 1 4 \int d^4 x \sqrt{-g} e^{-2\phi}  F_{\mu\nu} F^{\mu\nu} 
- \frac {\omega}2 \int d^4 x 
\sqrt{-g}  \pr_{\mu} \phi  \pr^{\mu} \phi,
\eea 
where electromagnetic field strength tensor can be expressed as
\bea
F_{\mu \nu }=(e^\phi {\bf a}_{\nu} )_{,\mu }-(e^\phi
{\bf a}_{\mu} )_{,\nu } = {\bf A}_{\nu,\mu }- {\bf A}_{\mu,\nu }. 
\eea
with ${\bf A}_{\mu} = e^{\phi} {\bf a}_{\mu}$ as the new 
electromagnetic gauge potential.

In the above action and for the rest of this paper we 
set $e_0=1$ for convenience.
As can be easily seen, the above action reduces to the usual 
form when $\phi$ is constant. The coupling constant $\omega$ is related to a 
characteristic mass scale of the theory above which the 
Coulomb force law is valid for a point charge. 
From the present experimental 
constraints the energy scale  has to be above a 
few tens of MeV to avoid conflict with experiments. 
At this point we to want mention that because of the underlying
shift symmetry, we cannot add any arbitrary potential in our Lagrangian. This
essentially says that the scalar field responsible for the variation 
of fine structure constant should be massless. Of course one can break this 
shift symmetry by introducing a potential term which has recently been studied
in \cite{bali}. We will keep this for our future study in the context of PCP violating
varying alpha theory.

One of the natural assumptions in constructing the above Lagrangian is 
time-reversal invariance. We have relaxed this 
assumption and try to analyse its implications based on 
various laboratory based experiments. 
An obvious term that is consistent with the varying alpha
framework yet violates PCP is ${\tilde F}_{\mu\nu} F^{\mu\nu}$, 
where ${\tilde F}^{\mu\nu} = \epsilon^{\mu\nu\sigma\rho} F_{\sigma\rho}$
is the Hodge dual of the Electromagnetic field tensor. 
In the conventional electromagnetism this does not 
contribute to the classical equation of motion.
But in the present framework this is no longer true because of its
coupling with the scalar field $\phi(x)$. As we have explained 
in the introduction, at the present level of experimental 
accuracy PCP violation in the electromagnetic sector may not be 
ruled out, and if the PCP in this EM sector is indeed violated, 
then there should have some interesting 
consequences. Motivated by this, we have introduced a parity violating
Lagrangian \cite{debupisin}
\be
{\cal L} ~=~  \frac {\omega} 2 \pr_{\mu}\phi 
\pr^{\mu} \phi ~+  \frac 1 4 e^{-2\phi } 
F_{\mu\nu} F^{\mu\nu} - \frac {\beta}{4}e^{-2\phi } 
F_{\mu\nu} {\tilde F}^{\mu\nu}~+~ 
{\mathcal L}_m,    \label{action}
\ee
where $\beta$ is
a free coupling parameter in our model.
As we can see, the scalar field $\phi$ plays a
similar role as that of the dilaton in the 
low-energy limit of string theory and M-theory, with the important 
difference that it induces a PCP violating electromagnetic 
interaction in our case. For our purpose, we assume $\beta$ as a free but small parameter.
Here we want to emphasize that the model can be thought of as
a unified framework for dealing with different phenomena. 
At the present level of experimental accuracy,  
investigations of parity or charge-parity violating
beyond-standard model may shed some new light 
about the fundamental laws of physics. 
With the interest of phenomenological impacts on various experimental 
observations, subsequently we will discuss about some consequences
of our model based on the laboratory experiments.   

The equations of motion are 
\bea
&&\frac 1 {\sqrt{-g}} \pr_{\mu}(\sqrt{-g}  F^{\mu\nu}) + \pr_{\mu}\phi
(- F^{\mu\nu} + \beta \tilde{F}^{\mu\nu}) =  0, \\
&& \frac 1 {\sqrt{-g}} \pr_{\mu}(\sqrt{-g} \phi) = \frac {e^{-2 \phi}}
{2 \omega}\left[- F_{\mu \nu }F^{\mu \nu} ~+~ 
\beta F_{\mu \nu } \tilde{F}^{\mu \nu} \right].  
\eea

We now explore the theoretical predictions of such coupling on 
the rotation of plane of polarization as well as ellipticity for an 
electromagnetic wave propagating through a transverse magnetic field.

\section{Calculation of Optical rotation and ellipticity} \label{sec2}
In this section we explicitly estimate the rotation angle of the 
plane of polarization and the
ellipticity due to the PCP violating scalar-photon 
coupling of our model.
The equations of motion for Maxwell and scalar fields turn out to be,
\bea
\nabla \cdot {\bf E} &=& 2 \nabla { \phi} \cdot {\bf E}- 
4 \beta \nabla { \phi} \cdot {\bf B} , \nonumber \\
\partial_{\eta}({\bf E}) - \nabla \times {\bf B} &=&
2 ({\dot { \phi}} {\bf E}-
\nabla { {\phi}}\times {\bf B}) -4 \beta( {\dot{{\phi}}}
{\bf B} + \nabla {{\phi}} \times {\bf E}),\nonumber \\
\bf \nabla \cdot \bf B &=& 0,\nonumber \\
\partial_{\eta}\bf B + \bf \nabla \times \bf E &=& 0.
\eea
As we have mentioned in the previous section,
the definition of electromagnetic field strengths
are $F_{i0}= (\pr_i {\bf A}_0 -\pr_0 {\bf A}_i)
= {\bf E}_i$ and $ F_{ij}= 
(\pr_i {\bf A}_j - \pr_j {\bf A}_i)=
\epsilon_{ijk} {\bf B}_k$, where $i=1,2,3$ and $\epsilon$
is the three spatial dimensional Levi-Civita tensor density.

At this point it is worthy of mentioning that different theoretical 
models based on the 
$e^{-2 \phi} F_{ab}{F}^{ab}$ type scalar-photon coupling or the standard QCD 
$\phi F_{ab}{\tilde F}^{ab}$ type axion-photon coupling 
mediated by the background magnetic or electric field have
been considered extensively 
\cite{raffelt,zavattini,pdpj,birch,kendall,cf,cfj,ringwald}. 
In a PCP violating varying alpha
theory we have both coupling terms. 
This motivates us to study the effect of 
both terms under the externally applied magnetic field.  

In terms of the vector potential ${\bf A}_{\mu}$ and the scalar 
field $\phi$, the linear order fluctuation equation in the
strong background magnetic field ${\bf B}_0$ takes the form
\bea \label{equations}
&&(\nabla^2  + \varpi^2) {\bf A}_x = 4 i \beta {\bf B}_0 \varpi \phi, \\
&&(\nabla^2 + \varpi^2) {\bf A}_y =- 2{\bf B}_0 \pr_z \phi, \\
&&(\nabla^2 + \varpi^2) {\bf A}_z = 2 {\bf B}_0 \pr_y \phi, \\
&&(\nabla^2 + \varpi^2) \phi = \frac {2 {\bf B}^2_0}{\omega} \phi 
- \frac {2 {\bf B}_0}{\omega}
(\pr_y A_z - \pr_z A_y) - 
\frac {4 i \beta {\bf B}_0 \varpi}{\omega} A_x .
\eea
We assume the background magnetic field ${\bf B}_0$ is applied
in the x-direction. 
In the above derivation we use the gauge condition 
$\nabla \cdot {\bf A} = 0$ and specify the scalar 
potential ${\bf A}_0 = 0$. $\varpi$ is the frequency of the 
electromagnetic field. Now, in general it is very difficult to 
solve the above equation. We will try to solve it by
choosing an appropriate ansatz for the electromagnetic field
which is usually used the laboratory set up. Therefore,
taking the propagation direction of the electromagnetic wave to be 
orthogonal to the external magnetic field ${\bf B}_0$ say z-direction, we
take the ansatz to be
\bea
{\bf A}(z,t) = {\bf A}^0 e^{- i \varpi t + i k z}~~~ ;~~~
 \phi(z,t)=\phi^0 e^{- i \varpi t + i k z}.  
\eea
As is clear from the above ansatz that
the equation for $A_z$ is no longer coupled
with $\phi$. So, evolution of this component of a vector 
potential does not get effected by the external magnetic field. 
The other three equations for $A_x,A_y,\phi$ turns out to be
\bea \label{mateq}
\begin{bmatrix} (\varpi^2-k^2) & 0 & - 4 i \beta {\bf B}_0 \varpi  \\
0 & (\varpi^2 -k^2) &  2 i {\bf B}_0 k  \\
\frac {4 i \beta {\bf B}_0 \varpi }{\omega} & 
- \frac {2 i {\bf B}_0 k }{\omega} & (\varpi^2 -k^2 -
\frac {2 {\bf B}_0^2} {\omega}) \\ \end{bmatrix}\left(\begin{array}{c}
 {\bf A}_x\\ {{\bf A}}_y\\ {\phi}\\
\end{array}\right)=0 ,
\eea
In order to have a consistent solution of the above matrix Eq.\ref{mateq},
the  determinant of this $3\times 3$ matrix part of this 
equation should be zero.
This consistency condition leads to three possible roots for the 
frequency $\varpi$ of a electromagnetic field as follows
\bea
\varpi^2  &=& k^2~~~,~~~\varpi_{\pm}^2 = k^2 + \delta_{\pm}\\
\delta_{\pm}&=& \frac {{\bf B}_0^2 } {\omega}(1 + 8 \beta^2)  \pm
\sqrt{\frac {{\bf B}_0^4 } {\omega^2}(1 + 8 \beta^2)^2 +
\frac {4{\bf B}_0^2 k^2}
{\omega} (1+4\beta^2)}.
\eea  
Now, to establish a connection with the experimental set up, 
we consider the initial(t = 0,x=0) electro-magnetic field to be 
linearly polarized and making an angle  with the external magnetic 
field ${\bf B}_0$ , so that 
\bea
{\bf A}_x(z=0,t=0) = \cos \alpha~~;~~  
{\bf A}_y (z=0,t=0)=  \sin\alpha ~~;~~ \phi(z=0,t=0)= 0.
\eea
With this boundary conditions, we have a unique solution for the above
system of equation as 
\bea
&&{\bf A}_x = (a_x e^{-i \varpi t} + b_x e^{-i \varpi_{+} t}+ 
c_x e^{-i \varpi_{-} t})e^{i k z} ,\nno\\
&&{\bf A}_y=( a_y e^{-i \varpi t} + b_y e^{-i \varpi_{+} t}+ 
c_y e^{-i \varpi_{-} t})e^{i k z},  \nno\\
&&\phi = \phi_0 ( e^{-i \varpi_{+} t} -  e^{-i \varpi_{-} t}) e^{i k z}  ,
\eea
where
\bea
b_x &=&- \frac {2 \beta \varpi_{+}}{k} b_y = \frac {2 \beta \varpi_{+}}{k} 
\frac {\delta_-}{\delta_+} c_y = -\frac {\varpi_{+}}
{\varpi_{-}} \frac {\delta_-}{\delta_+} c_x 
= \frac {4 i \beta {\bf B}_0 \varpi_{+}}{\delta_+} \phi_0,
\nno\\
a_y &=& \frac {2 \beta \varpi}{k} a_x = \sin \alpha + \frac{k} {2 \beta 
\varpi_- } 
\left(\frac {\delta_{+}-\delta_{-}}{\delta_+}
\right) c_x ,\nno \\
c_x &=& \frac {1}{{\cal F}}\left(\cos \alpha - \frac {k \sin \alpha}
{2 \beta \varpi}\right) ,\nno \\
{\cal F}&=& \frac {4 \beta^2 (\varpi \varpi_{-} 
\delta_+-\varpi \varpi_{+}
\delta_-)+ k^2(\delta_+-
\delta_-)}{4 \beta^2 \varpi \varpi_{-}\delta_+}. 
\eea
After traversing the external magnetic field for a distance $t=\ell$,
the electromagnetic wave will be modified 
as
\bea \label{soll}
&&{\bf A}_x = a_x e^{-i \varpi \ell} + b_x e^{-i \varpi_{+} \ell}+
c_x e^{-i \varpi_{-} \ell}, \nno \\
&&{\bf A}_y= a_y e^{-i \varpi \ell} + b_y e^{-i \varpi_{+} \ell}+
c_y e^{-i \varpi_{-} \ell}.
\eea
From the above set of expressions, one can easily see
that the vector potential describes an ellipse whose major
axis deviates from the $x$-axis by an angle
\bea \label{rot}
\theta = \tan^{-1}\left( \sqrt{\frac {\sin^2 \alpha - {\Gamma}} 
{\cos^2 \alpha - {\cal L}}}\right),  
\eea
where
\bea
&&{\cal L} = 2 a_x b_x \sin^2(\frac {\Delta_{+}}{2}) + 2 a_x c_x 
\sin^2(\frac {\Delta_{-}}{2})+2 c_x b_x \sin^2(\frac {\Delta}{2}), \nno\\
&&{\Gamma} = 2 a_y b_y \sin^2(\frac {\Delta_{+}}{2}) + 2 a_y c_y 
\sin^2(\frac {\Delta_{-}}{2})+2 c_y b_y \sin^2(\frac {\Delta}{2}), \nno\\
&&\Delta_{+} =( \varpi_{+} -\varpi) \ell~~~;~~~\Delta_{-} = 
(\varpi_{-} -\varpi) \ell~~~;~~~
\Delta = (\varpi_{+} -\varpi_{-}) .\ell
\eea
Eq.(\ref{rot}) yields the expression for the 
optical rotation of the plane of polarization as
\bea \label{rotation}
\delta = \theta -\alpha \simeq 
\frac {\sin (2 \alpha)} {4}\left( \frac {{\cal L}}
{\cos^2(\alpha)} - \frac {{\Gamma}} {\sin^2(\alpha)} \right).
\eea

The ellipticity $\epsilon$ is the measure of the phase difference 
between the two components of the vector potential
after traversing a distance $\ell$ through a magnetic region.
From Eq.(\ref{soll}) the exact expression for the 
ellipticity is
\bea \label{ellipticity}
\epsilon =  \frac 1 2 |\psi_x -\psi_y|, \nno
\eea
where
\bea 
\psi_x &=& \tan^{-1}\left[\frac{bx \sin(\Delta_{+}) + cx \sin (\Delta_{-})}
{ax + bx \cos(\Delta_{+} ) + cx \cos(\Delta_{-}l)} \right] ,\nno\\
\psi_y &=& \tan^{-1}\left[\frac{by \sin(\Delta_{+}) + cy \sin (\Delta_{-})}
{ay + by \cos(\Delta_{+} ) + cy \cos(\Delta_{-})} \right] .
\eea

These are the quantities that establish the direct connection with the 
experimental data. The similar analysis can be done for the background
electric field as well. The analysis we have done so far
is applicable to the laboratory experimental observations where a polarized monochromatic laser beam with a fixed momentum traverses a magnetic region. If we want instead to consider
an unpolarized light, then we have to solve the above coupled system of
Eqs.(\ref{equations}) in term of the spacetime coordinates. 
For example in CMB polarization power 
spectrum, the initial state of the electromagnetic wave
at the last scattering surface is completely unpolarized. So our
present analysis in not adequate to study CMB polarization. 
In our forthcoming paper we will consider more
detail analysis of the background electromagnetic field effect 
on the scalar-photon mixing and will study its
cosmological connection.
In the next section we will consider various existing laboratory
based experimental results on the optical rotation $\delta$ 
and ellipticity $\epsilon$ to constrain our model parameter
$\beta$ and scale of varying fine structure constant $\omega$.

\section{Constraining $\beta$ and $\omega$ parameters
through laboratory experiments} \label{sec3}
The polarization properties of an electromagnetic wave propagating
through an external magnetic field can change if there
exist a non-trivial (pseudo)scalar-photon coupling \cite{zavattini}. 
Based on this particular physical effects, various laboratory 
based experiments have been devised to look for 
ultra-light (pseudo)scalar particles. 
In this class of experiments, it is possible to 
make accurate measurements on the
modification of the polarization state of a light beam. In a typical
experiment, a linearly polarized 
laser beam is used to reflect N times between two
mirrors, in a constant strong background
magnetic field of strength ${\bf B}_0$.
The magnetic field is perpendicular to the beam direction. 
Let the distance between the two mirrors  
be $\ell$, then the total length travelled by the
laser beam in the magnetic field is $L = N\ell$. 
After traversing a distance $L$, which is usually
of the order of a few Kilometers, it is possible to measure a minute
ellipticity and a change in the rotation of the polarization plane.

As is well known, the vacuum itself has the 
magnetic birefringence property as dictated by QED. This effect is due to
the dispersive effect induced by the virtual electron-positron pair in vacuum, which was first investigated by Heisenberg and Euler \cite{heisenberg}. The ellipticity so induced
serves as the background in the experiment that looks for
birefringence or dichroism induced by a (pseudo)scalar particle that violates PCP. The QED
contribution to the ellipticity can be written as

\begin{eqnarray}
{\cal E} = N \frac {B^2_0 \ell \alpha_0^2 \omega} {15 m_e^4},
\label{qed}
\end{eqnarray}
where $\alpha_0 =e_0^2= 1/137$ is the conventional fine-structure constant, $\omega$ is the photon
energy and $m_e$ the electron mass. Here we have assumed that the polarization
vector of the initially linearly polarized beam makes a 45$^\circ$ angle relative to
the direction of the external magnetic field. Consider, for example, a laser beam with wavelength $\lambda$ = 1550 nm, magnetic field $B_0$ = 9.5 T, and length of travel $N\ell$ = 25 km. Then the
resulting ellipticity ({\it cf.} Eq.(\ref{qed})) would be $2 \times 10^{-11}$ rad
\cite{battesti}.

It is important to note that any physical mirror is transparent
to scalar field so that only the photon component of the beam is reflected. 
As in varying alpha theory, matter field $\psi$ coupled with 
the scalar field $\phi$ through its electromagnetic 
mass correction at the loop level:
\bea
{\cal L}_{int} \sim \frac {e_0^2} {\omega}  {\bar \phi} 
m_{\psi}^2 {\bar \psi} \psi + 
e_0 A_{\mu} \gamma^{\mu} {\bar \psi} \psi,
\eea
where, we consider ${\bar \phi}$ as the dimensionfull scalar field. 
It is natural to expect and obvious from the above equation that
compared with the photon coupling, $\phi$ coupling is significantly
suppressed.  
This essentially sets the scalar component of the beam back to zero after each
reflection \cite{raffelt}. The net effect after N reflections is
${\cal E}(L) = N {\cal E}(\ell)$ where, in general, $N {\cal E}(\ell)$ does not equal to ${\cal E}(N\ell)$.
Thus, in order to take into account the effect of N reflections appropriately
for a multiple-beam-path experiment, one needs to multiply the right-hand side
of Eq.(\ref{rotation}) and Eq.(\ref{ellipticity}) 
by N (keeping everything else the same)
while on the left-hand side, the length $\ell$ of a single-path is now replaced by
the total length $L = N \ell$.

\begin{table}[t!]
\begin{tabular}{|c|c|c|c|c|c|c|}
\hline
Experiment & $\lambda (nm)$ &${\bf B}_0 (T)$ & $L(m)$ & $N$&
Rotation $\delta$ (rad) & Ellipticity $\epsilon$ (rad)\\
\hline
BFRT& 514 & 3.25&8.8&250 &$3.5\times 10^{-10}$ & -\\
\hline
PVLAS& 1064& \begin{tabular}{c} 2.3\\ 5.5\end{tabular} 
& 1& 45000& \begin{tabular}{c} $1.0 \times 10^{-9}$ \\ $1.2 \times 10^{-8}$
\end{tabular} &\begin{tabular}{c}$1.4 \times 10^{-8}$\\-\end{tabular} \\
\hline
Q\&A & 1064&2.3&0.6&18700&$(-0.375\pm 5.236)\times10^{-9}$& -\\
\hline
\end{tabular}
\caption{Laboratory 
experiments and their bounds on rotation $\delta$ and
ellipticity $\epsilon$ measurements} \label{tab1}
\end{table}

In our analysis we will consider various laboratory 
experiments as listed in the Table \ref{tab1}. In all these
experiments, a polarized
laser beam with a particular wavelength has been used
to measure the rotation and ellipticity of polarization. As we have
discussed before, the basic underlying 
assumption of all these measurements is that there
exists a non-trivial interaction of photon with the 
background field. The nature of this background field
could be a Lorentz invariant (pseudo) scalar or
some Lorentz violating vector field. The scalar 
field could be a QCD axion or some arbitrary dilaton field.
As we saw, the scalar field that gives rise to the spacetime variation of fine structure constant
can also induce
the polarization of an electromagnetic wave through its
PCP violating interactions. Rotation or ellipticity measurement
therefore cannot distinguish different nature of the 
background interaction.
Evidently all the above experiments we mentioned are
insensitive to the nature of the background field. 
Therefore by analyzing the above experimental
bounds on optical rotation $\delta$ and ellipticity $\epsilon$
one can only constrain the parameters of a particular
model without the explicit nature of the background field. 
It is therefore interesting to find out 
an observable that can distinguish different models. 

All the above mentioned 
experiments have reported the upper limit on the optical
rotation $\delta$ except that PVLAS has reported the upper limit on 
ellipticity as well. It should be emphasized that 
in the case of the Q\&A experiment, the error bar in the optical 
rotation measurement is very large compared to its mean value.
The amount of the optical rotation would therefore not be very conclusive but
we will still use its mean value for its upper bound to constrain our model 
parameters and compare with the other experimental constraints.    

In Fig.1 we have plotted the shaded exclusion regions in the two-dimensional
plane spanned by the PCP violating parameter $\beta$ and
the scale of variation of the fine structure constant $\omega$.
Each plot corresponds to different experimental results.
From these plots we see that the bound on the scale 
of fine structure constant variation based on the absence of ellipticity measured by PVLAS is $\omega \gsim 10^{-10}~{\rm GeV}^2$. 
On the other hand, the bound on the PCP violating parameter $\beta$
based on the optical rotation measurement is  
almost the same for every experiments. 
As we can see from the contour plots of these experiments, 
the bound on PCP violating parameter
falls into three different ranges as shown in Table \ref{tab2}.

\begin{figure}[t!]
\includegraphics[width=6.3in,height=2.5in]{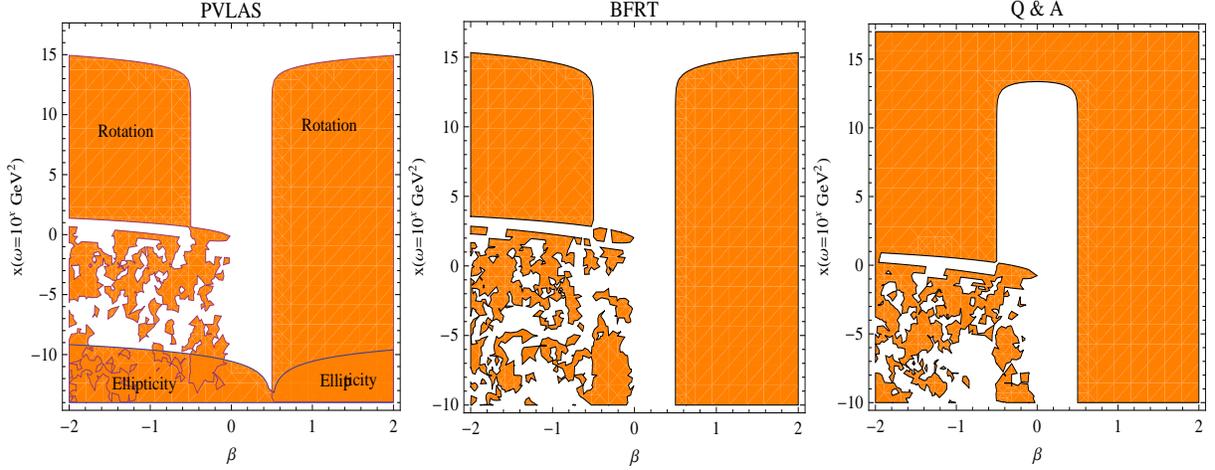}
\caption{\label{fig1} Bounds on PCP violating parameter 
$\beta$ and varying alpha theory mass scale square $\omega$ 
using different experimental results for the rotation and the 
ellipticity}
\end{figure}

\begin{table}[t]
\begin{tabular}{|c|c|c|}
\hline
Range of $\omega$ in $\mbox{GeV}^2$ & Bound on $\beta$& Experiment\\
\hline
\begin{tabular}{r}$10^{-10} \lessapprox \omega \lessapprox 10^2$\\
 $\omega \lessapprox 10^2$\end{tabular} & $0 \leq \beta \leq 0.5$ & 
\begin{tabular}{c} PVLAS \\ BFRT, Q$\&$A \end{tabular} \\
\hline
$10^{2} \lessapprox \omega \lessapprox 10^{13}$ & $-0.5 \leq \beta \leq 0.5$ & 
PVLAS, BFRT, Q$\&$A \\
\hline
$\omega \gtrapprox 10^{14}$ & $ |\beta| \geq 1$ & PVLAS, BFRT \\
\hline
\end{tabular}
\caption{Laboratory constraints on PCP violating parameter $\beta$} 
\label{tab2}
\end{table}

In the above analysis we have excluded the fluctuating region 
of $\omega \leq 1~ \mbox{GeV}^2$ and $\beta <0$. For this
range of $\omega$, the bound on PCP violating parameter 
becomes $0 <\beta< 0.5$ as shown in the table \ref{tab2}.
Interestingly, if we consider the negative mean value of
the optical rotation $\delta = -0.375$ based on the Q$\&$A experiment, 
the contour plot shows that $\omega(\geq  10^{13}~ \mbox{GeV}^2) $ is 
bounded from above. This in turn constrains the PCP violating 
parameter $\beta$ to be always
less than one. The main concern, however, is that 
the error-bar in this particular observational constraint
on the optical rotation is much larger than its mean value.
So this upper bound on $\omega$ may not be conclusive.

We have mentioned before that in order to be consistent
with the observation of Coulomb force law, 
the scale of fine structure constant 
variation ${\sqrt{\omega}}={\hbar}c/l$ should be 
greater than a few tens of $\mbox{MeV}$. 
From the ellipticity measurement of PVLAS, 
we found the lower bound on this scale to be $\approx 10^{-2}$ MeV, which is
in direct conflict with Coulomb force law measurement.
According to our model, if we assume a lower bound on 
${\sqrt{\omega}} \approx 10^3$ MeV, then with the present value of 
the experimental input parameters, PVLAS would not 
be able to measure the ellipticity down to the level of 
$\epsilon \simeq 1 \times 10^{-17}$. 
This value is significantly lower than the present 
PVLAS bound of $\epsilon \simeq 1.4 \times 10^{-8}$.  
One can also see from Table \ref{tab2} that
for $\omega \gtrapprox 10^{14}$ $\mbox{GeV}^2$, 
the bound is $|\beta|\geq 1$ according to PVLAS and BFRT.
This should be unacceptable in connection with
the other physical parity violating effects due to large PCP coupling.
From all the above considerations, we conclude that the most reasonable bounds on 
both of our model parameters are
$1 \leq \omega \leq 10^{13}~ \mbox{GeV}^2$ and
$-0.5 \leq \beta \leq 0.5$.

\section{Conclusions} \label{con}
The theory of varying fine structure constant has been the 
subject of intense study
for the last several years. Cosmological impact of this
variation has been studied quite extensively. Various cosmological
as well as laboratory based observations on this variation 
of fine structure constant have been considered to constrain
the varying alpha parameter
$\omega$. Recently we have constructed a particular model based on 
this varying alpha theory which includes explicit PCP
violation in the photon sector \cite{debupisin}.  
In this paper we have studied our aforementioned PCP violating 
varying alpha model in the light of a new class of 
laboratory observations which have 
not been considered before. 
All those experiments directly measure the 
change of the polarization state of a photon as it propagates 
through a background magnetic field. 
The basic underlying assumption behind all these measurements is
the existence of a non-trivial interaction between photon and 
some unknown background field. As stated before in our model
we have introduced a non-trivial PCV violating scalar-photon 
interaction in the varying alpha theory framework. Although 
the experiments under consideration are
insensitive to the properties of the background field 
due to the weakness of its coupling with the matter, 
they nevertheless can help to constrain our varying alpha model parameters 
$\omega$ and $\beta$ through the rotation of polarization and ellipticity 
measurement. We have calculated these two particular 
measurable quantities in our model.
The model is characterized by two independent parameters
$\beta$ and $\omega$ that measure the strength of 
PCP violation and the scale of fine structure constant variation, 
respectively. We have considered three different laboratory
based experimental results to constrain our model parameters. 
All the experiments so far do not observe any positive signal for
the rotation and ellipticity. So, the non-observation 
give us the possible upper limit on those quantities.  
Using those upper limits, we found that the most suitable bound
on our two model parameters are $1 \leq \omega \leq 10^{13} \mbox{GeV}^2$ and
$-0.5 \leq \beta \leq 0.5$. An interesting point to note here is that
with this lower bound on $\omega$, it is very hard to measure
the ellipticity from laboratory experiments. As we have estimated,
for $\omega \simeq 10^3$ MeV, the bound on ellipticity should be
$\epsilon \simeq 1\times 10^{-17}$,
which is far bellow the present experimental limit as well as 
sensitivity.

\vspace{.1cm}

{\bf Acknowledgement}\\
This research is supported by Taiwan National Science Council under Project No. NSC
97-2112-M-002-026-MY3, by Taiwan's National Center
for Theoretical Sciences (NCTS), and by US Department
of Energy under Contract No. DE-AC03-76SF00515.

\end{document}